\documentclass{aastex631}

\hypersetup{linkcolor=red,citecolor=blue,filecolor=cyan,urlcolor=magenta}

\def\lessim{\mathrel{\hbox{\rlap{\hbox{\lower4pt\hbox{$\sim$}}}\hbox{$<$}}}}
\def\grtsim{\mathrel{\hbox{\rlap{\hbox{\lower4pt\hbox{$\sim$}}}\hbox{$>$}}}}

\usepackage{CJK}

\shorttitle{M31N 2013-10c}
\shortauthors{Shafter et al.}

\graphicspath{{./}{figures/}}

\begin{document}
\begin{CJK*}{UTF8}{gbsn}
\title{M31N 2013-10c: A Newly Identified Recurrent Nova in M31}

\correspondingauthor{A. W. Shafter}
\email{ashafter@sdsu.edu}

\author[0000-0002-1276-1486]{Allen W. Shafter}
\affiliation{Department of Astronomy, San Diego State University, San Diego, CA 92182, USA}

\author[0000-0002-0835-225X]{Kamil Hornoch}
\affiliation{Astronomical Institute of the Czech Academy of Sciences,
Ond\v{r}ejov, Czech Republic}

\author[0000-0002-1330-1318]{Hana Ku\v{c}\'akov\'a}
\affiliation{Astronomical Institute, Charles University, Faculty of Mathematics and Physics,
Prague, Czech Republic}

\author[0000-0003-1127-9302]{Petr Fatka}
\affiliation{Astronomical Institute of the Czech Academy of Sciences,
Ond\v{r}ejov, Czech Republic}

\author[0000-0002-2770-3481]{Jingyuan Zhao (赵经远)}
\affiliation{Xingming Observatory, Mount Nanshan, Xinjiang, China}

\author[0000-0002-7292-3109]{Xing Gao (高兴)}
\affiliation{Xingming Observatory, Mount Nanshan, Xinjiang, China}

\author{Shahidin Yaqup (夏伊丁·亚库普)}
\affiliation{Xinjiang Astronomical Observatory, Chinese Academy of Sciences, Xinjiang, China}

\author{Tuhong Zhong (钟土红)}
\affiliation{Xinjiang Astronomical Observatory, Chinese Academy of Sciences, Xinjiang, China}

\author[0000-0003-1845-4900]{Ali Esamdin (艾力·伊沙木丁)}
\affiliation{Xinjiang Astronomical Observatory, Chinese Academy of Sciences, Xinjiang, China}

\author{Chunhai Bai (白春海)}
\affiliation{Xinjiang Astronomical Observatory, Chinese Academy of Sciences, Xinjiang, China}

\author[0000-0002-9786-8548]{Na Wang (王娜)}
\affiliation{Xinjiang Astronomical Observatory, Chinese Academy of Sciences, Xinjiang, China}

\author[0000-0001-6981-8722]{Paul Benni}
\affiliation{Acton Sky Portal Private Observatory, Acton, MA, USA}

\author{Aiden Luo}
\affiliation{Westview High School, San Diego, CA 92129, USA}

\author{Ilana Yousuf}
\affiliation{Westview High School, San Diego, CA 92129, USA}

\begin{abstract}

The nova M31N 2023-11f (2023yoa) has been recently identified as the second eruption of a previously recognized nova,
M31N 2013-10c, establishing the latter object as the 21st recurrent nova system thus far identified in M31. Here we present
well sampled $R$-band lightcurves of both the 2013 and 2023 eruptions of this system. The
photometric evolution of each eruption was quite similar as expected for the same progenitor system.
The 2013 and 2023 eruptions each reached peak magnitudes just brighter than $R\sim16$, with
fits to the declining branches of the eruptions yielding times to decline by two magnitudes
of $t_2(R)=5.5\pm1.7$ and $t_2(R)=3.4\pm1.5$ days, respectively. M31N 2013-10c
has an absolute magnitude at peak, $M_R=-8.8\pm0.2$, making it the most luminous known recurrent
nova in M31.

\end{abstract}


\section{Introduction} \label{sec:intro}

A recent nova in the Andromeda galaxy, M31N 2023-11f, has been confirmed by \citet{2023ATel16354....1S} to be the second known eruption of a previously identified nova M31N~2013-10c \citep[][]{2013ATel.5468....1H,2013ATel.5503....1H}.
The identification of M31N 2013-10c as a recurrent nova (RN) brings to 21 the total number of such systems known in M31 -- more than
twice the number of RNe known in the Milky Way \citep[][]{2010ApJS..187..275S}. The observed time interval
between eruptions establishes an upper limit of $\sim$10.1~yr on the recurrence time (it is conceivable, though unlikely, that one or more eruptions may have been missed). Such a short
recurrence time is typical of RNe in M31, but is quite short by Galactic standards where of the 10 RNe known, only U Sco ($t_\mathrm{recur} \simeq 10.3$~yr)
has a comparably short recurrence interval.
Here, we present $R$-band lightcurves for the two known eruptions of M31N 2013-10c, demonstrating that, as expected, they are quite similar. 

\begin{figure*}
\includegraphics[angle=0,scale=0.72]{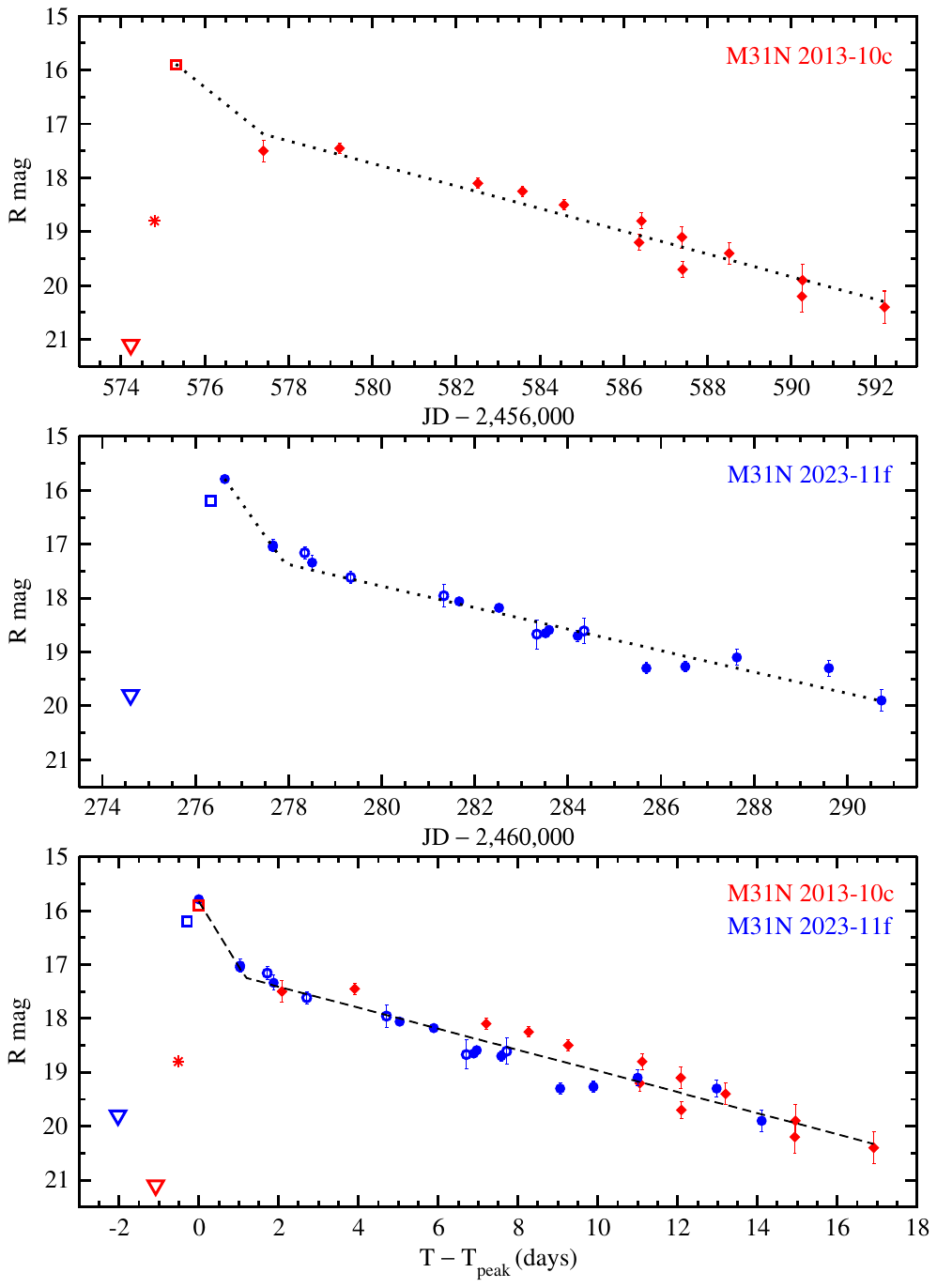}
\caption{The lightcurves for the 2013 October (M31N 2013-10c, top panel) and 2023 November (M31N 2023-11f, middle panel) eruptions of the RN M31N 2013-10c. The bottom panel shows both eruptions superimposed after adopting times of maximum of JD 2,456,575.313 and 2,460,276.630 for the 2013 and 2023 eruptions, respectively. The star and the open square in the upper panel corresponds to the photometry reported by the LOSS and N. James, respectively, while the open square in the middle panel shows the unfiltered discovery magnitude of \citet{2023TNSTR3081....1X}. The open circles in the middle panel represent Sloan $r$ photometry converted to Cousins $R$ as described in the text. The downward-facing triangles represent lower limits from prediscovery images.  The dashed line in the lower panel represents the best-fitting two-component line used to estimate $\langle t_2(R) \rangle = 3.9\pm1.0$~d. The photometric data are given in Table~A1.
}
\label{fig:f1}
\end{figure*}

\section{The Light Curves}

M31N 2013-10c was discovered by the Lick Observatory Supernova Search (LOSS) on 2013 October 09.30 UT at a magnitude, $m=18.8$ (unfiltered). Shortly thereafter, an observation by
N. James on October 09.813 UT revealed that the object had brightened to $m=15.9$ (unfiltered)\footnote{\tt http://www.cbat.eps.harvard.edu/unconf/followups/J00430954+4115399.html}. A series of $R$-band measurements were subsequently
made with the 0.65-m reflector at the Ond\v{r}ejov Observatory.
The top panel of Figure~\ref{fig:f1} shows the full lightcurve for M31N 2013-10c.
The declining portion of the light curve for both eruptions is best represented by a two-component linear fit (shown as broken lines in Figure~\ref{fig:f1}).
The fit to the M31N 2013-10c data suggests $t_2(R)=5.5\pm1.7$~d.

M31N 2023-11f was discovered by \citet{2023TNSTR3081....1X} at $m=16.2$ (unfiltered) on 2023 November 27.831 UT. A subsequent measurement
made on 2023 November 28.130 UT showed that the nova had brightened to $R=15.79\pm0.04$, which we take as maximum light. A series of
$R$-band measurements over the next 2 weeks were made at the Mount Laguna, Ond\v{r}ejov, and La Silla observatories. These observations were supplemented by Sloan $r$ measurements taken at the Xinjiang Observatory, which have been converted to $R$ using the transformations given in \citet{2006A&A...460..339J}. The resulting
lightcurve is presented in the middle panel of Figure~\ref{fig:f1}. The corresponding fit to the declining portion of the lightcurve
(dotted line) yields $t_2(R)=3.4\pm1.5$~d.

The bottom panel shows the two eruptions superimposed with the abscissa showing the time relative to that of
peak brightness for each eruption. It is clear that the lightcurves of the two eruptions are similar, with peak observed magnitudes of
$m=15.9$ (unfiltered) and $R=15.79$ for the 2013 and 2023 eruptions, respectively, and rates of decline that are also similar.
A two-component fit of the combined lightcurve data yields a mean time to decline 2 mag from maximum light, $\langle t_2(R) \rangle =3.9\pm1.0$~d.

At the distance of M31 \citep[$\mu_0=24.42\pm0.06$,][]{2013AJ....146...86T}, and adopting a foreground $R$-band extinction of 0.14 mag \citep[][]{2011ApJ...737..103S}, we find that M31N 2013-10c reaches an absolute magnitude, $M_R\simeq-8.8\pm0.2$, where we have estimated a generous uncertainty taking into account the possibility that maximum light might have been missed.

\section{Discussion}

Models show that a relatively high rate of accretion ($\dot M \grtsim 10^{-7}$~M$_\odot$~yr$^{-1}$)
onto a massive white dwarf ($M_{WD} \grtsim 1.3$~M$_\odot$) produce relatively weak nova eruptions that recur frequently ($t_\mathrm{recur} \lessim 100$~yr) an eject a relatively small amount of mass \citep[e.g., see][]{2014ApJ...793..136K}. The small ejected mass results in a relatively rapid photometric evolution, with $t_2$ times typically under 10 days. These properties result in an M31 RN population that is characterized by the following median values $\langle t_\mathrm{recur} \rangle = 9.6$~yr, $\langle M_R \rangle = -7.0$, and $\langle t_2 \rangle = 6.0$~d. Clearly, M31N 2013-10c is not atypical in terms of its recurrence time and rate of decline, but is significantly more luminous
than the typical M31 RN system. 

Ongoing monitoring of M31 will confirm whether the recurrence time of M31N 2013-10c is indeed $\sim10$~yr, or whether it might be an
integer fraction of that value. Finally, 10 of the 13 M31 RNe that have been spectroscopically confirmed are members of the He/N class \citep[e.g., see][for a discussion of spectroscopic classifications]{1992AJ....104..725W}. Thus, it will also be important to obtain spectroscopy during the next eruption to establish the spectroscopic class of this unusually luminous RN.

\begin{acknowledgments}
We thank Mi Zhang (张宓),
Guoyou Sun (孙国佑), Wenjie Zhou (周文杰), and Jianlin Xu (徐建林)
for alerting us that M31N 2023-11f could be another outburst of 2013-10c.
P. Benni acknowledges the AAVSO.
\end{acknowledgments}

\vspace{5mm}
\facilities{Mount Laguna Observatory 40-in; Ond\v{r}ejov Observatory 0.65-m; Danish 1.54-m, La Silla Observatory; Xinjiang Astronomical Observatory 17-in Photometric Auxiliary Telescope; Acton Sky Portal 0.28-m SCT.}

\bibliography{M31N2013-10c}{}
\bibliographystyle{aasjournal}


\begin{deluxetable}{cccr}
\tabletypesize{\scriptsize}
\tablenum{A1}
\tablecolumns{4}
\tablecaption{Nova Photometry}
\tablehead{\colhead{($\mathrm{JD} - 2,400,000$)} & \colhead{Filter\tablenotemark{a}} & \colhead{Mag} & \colhead{Telescope\tablenotemark{b}}
}
\startdata
\cutinhead{M31N 2013-10c}
56574.234 &R &$>21.1$       &D65 \cr
56574.800 &C &$18.8$        &KAIT \cr
56575.313 &C &$15.90\pm0.10$&C11 \cr
56577.399 &R &$17.50\pm0.20$&D65 \cr
56579.218 &R &$17.45\pm0.10$&D65 \cr
56582.516 &R &$18.10\pm0.10$&D65 \cr
56583.582 &R &$18.25\pm0.10$&D65 \cr
56584.571 &R &$18.50\pm0.10$&D65 \cr
56586.365 &R &$19.20\pm0.15$&D65 \cr
56586.424 &R &$18.80\pm0.15$&D65 \cr
56587.390 &R &$19.10\pm0.20$&D65 \cr
56587.406 &R &$19.70\pm0.15$&D65 \cr
56588.516 &R &$19.40\pm0.20$&D65 \cr
56590.251 &R &$20.20\pm0.30$&D65 \cr
56590.265 &R &$19.90\pm0.30$&D65 \cr
56592.227 &R &$20.40\pm0.30$&D65 \cr
\cutinhead{M31N 2023-11f}
60274.605 &R &$>19.8$       &D65 \cr
60276.331 &C &$16.20\pm0.10$&XO50 \cr
60276.630 &R &$15.79\pm0.04$&SCT \cr
60277.661 &R &$17.05\pm0.06$&MLO \cr
60277.665 &R &$17.02\pm0.12$&SCT \cr
60278.347 &Rr&$17.16\pm0.12$&PAT \cr
60278.506 &R &$17.34\pm0.14$&SCT \cr
60279.335 &Rr&$17.62\pm0.12$&PAT \cr
60281.335 &Rr&$17.96\pm0.21$&PAT \cr
60281.663 &R &$18.06\pm0.05$&MLO \cr
60282.520 &R &$18.18\pm0.04$&D154 \cr
60283.334 &Rr&$18.67\pm0.27$&PAT \cr
60283.521 &R &$18.65\pm0.07$&D154 \cr
60283.596 &R &$18.59\pm0.08$&MLO \cr
60284.211 &R &$18.70\pm0.10$&D65 \cr
60284.347 &Rr&$18.61\pm0.24$&PAT \cr
60285.686 &R &$19.30\pm0.10$&MLO \cr
60286.521 &R &$19.27\pm0.10$&D154 \cr
60287.629 &R &$19.10\pm0.15$&MLO \cr
60289.608 &R &$19.30\pm0.15$&MLO \cr
60290.737 &R &$19.90\pm0.20$&MLO \cr
\enddata
\tablenotetext{a}{R -- Cousins $R$; Rr -- Sloan r converted to Cousins R; C -- Clear (no filter)}
\tablenotetext{b}{D65: Ond\^{r}ejov 0.65-m; KAIT: Katzman Automatic
          Imaging Telescope; C11: Celestron 11-in; XO50: Xingming 50-cm;
          SCT: Acton Sky Portal 28-cm Schmidt-Cassegrain;
          PAT: 17-in Photometric Auxiliary Telescope,
          Xinjiang Astronomical Observatory; MLO: Mount Laguna 40-in;
          DK154 - ESO 1.54-m Danish Telescope, La Silla.}
\end{deluxetable}
\end{CJK*}
\end{document}